\documentclass[twocolumn,showpacs,preprintnumbers,amsmath,amssymb]{revtex4}

\usepackage{graphicx}
\usepackage{dcolumn}
\usepackage{bm}

\begin{document}
\draft
\def\>{\rangle}
\def\w{\omega}
\def\W{\Omega}
\def\lr{\leftrightarrow}
\def\ud{\updownarrow}

\title{Entanglement and spin squeezing properties for three bosons in two modes}
\author{B. Zeng,$^1$ D. L. Zhou,$^{2}$ Z. Xu,$^{3}$ and L. You$^{2,4}$}

\address{$^1$Department of Physics, Massachusetts Institute of
Technology, MA 02139, USA}
\address{$^2$School of Physics,
Georgia Institute of Technology, Atlanta, Georgia 30332, USA}
\address{$^3$Center for Advanced Study, Tsinghua University, Beijing 10084, China}
\address{$^4$Interdisciplinary Center of Theoretical Studies, The Chinese
Academy of Sciences, Beijing 10080, China}
\date{\today}

\begin{abstract}
We discuss the canonical form for a pure state of three identical
bosons in two modes, and classify its entanglement correlation
into two types, the analogous GHZ and the W types as well known in
a system of three distinguishable qubits. We have performed a
detailed study of two important entanglement measures for such a
system, the concurrence $\mathcal{C}$ and the triple entanglement
measure $\tau$. We have also calculated explicitly the spin
squeezing parameter $\xi$ and the result shows that the W state is
the most ``anti-squeezing'' state, for which the spin squeezing
parameter cannot be regarded as an entanglement measure.
\end{abstract}

 \maketitle

\section{Introduction}
Quantum entanglement is an intriguing property of composite
systems. It refers to the inseparable correlations stronger than
all classical counterparts. Recent studies indicate that
entanglement is not only of interest to the interpretation of the
foundations of quantum mechanics, but also represents a useful
resource for quantum computation and quantum communication.
Inseparable correlations such as entanglement also exist in
systems of identical particles, e.g. electrons in quantum dots
\cite{slm}, atoms in a Bose-Einstein condensate \cite{You,slcp},
and electrons in quantum Hall liquids \cite{zzx}. Even for the
widely used parametric down conversion process, a complete
treatment must take into account the indistinguishability of the
down converted photons. Although the study of entanglement has had
a long history in systems of distinguishable particles, only
recently did the entanglement properties in a system of identical
particles begin to attract much attention
\cite{slm,sckll,py,lbll,gf,esbl,wiseman}.

Quantum correlation among identical particles was noted by
Schliemann {\it et al.} \cite{slm,sckll} and they discussed the
entanglement in a two-fermion system. It was argued that the
separability of a two fermion state should be defined in terms of
whether or not that state can be expressed in terms of a sum of
single Slater determinants \cite{slm}. More generally, a two
fermion pure state can always be expressed in the following
standard form \cite{sckll}
\begin{eqnarray}
{|\Psi\rangle=\frac{1}{\sqrt{\sum_{i=1}^{k}|z_{i}|^{2}}}
\sum\limits_{i=1}^{k}z_{i}f^{\dag}_{a_{1}(i)}f^{\dag}_{a_{2}(i)}|0\rangle},
\end{eqnarray}
where $f^{\dag}_{a_{1}(i)}|0\rangle$ and
$f^{\dag}_{a_{2}(i)}|0\rangle$ represent the orthonormal basis
states of a single particle (fermion).

The case of two bosons were considered independently by
Pa\v{s}kauskas and You \cite{py} and Li {\it et al.} \cite{lbll}.
They found a similar standard form for two-identical bosons,
namely, the wave functions of two bosons can always be written in
the following form
\begin{eqnarray}
{|\Psi\rangle=\sum\limits_{i=1}^{M}\lambda_{i}a_i^{\dag}a_i^{\dag}|0\rangle},
\end{eqnarray}
where $a_i^{\dag}|0\rangle$ forms an orthonormal basis in the
single particle (boson) space.

The above two results are in fact simple extensions of the well
known result for two distinguishable particles; that an arbitrary
pure state can be described in terms of the famous Schmidt
decomposition
\begin{eqnarray}
|\psi\rangle=\sum\limits_i\sqrt{\lambda_i}|i_1i_2\rangle ,
\end{eqnarray}
with real parameter $\lambda_i$ satisfying $\sum_i\lambda_i=1$
and $\langle i_m|j_n\rangle=\delta_{mn}\delta_{ij}$.
 A natural generalization of the Schmidt
decomposition to $N>2$ particles is
\begin{eqnarray}
|\psi\rangle=\sum\limits_i\sqrt{\lambda_i}|i_1 i_2\cdots
i_N\rangle \label{sd},
\end{eqnarray}
with the sub-indices $m$ ($n$) for the $m$-th ($n$-th) particle,
and $i$ ($j$) the $i$-th ($j$-th) basis vector. However, even for
three two-state particles or three qubits with the Hilbert space
${\cal H}=C^2\otimes C^2\otimes C^2$, the above Schmidt
decomposition does not exist, pointing to a truly challenging
prospect for characterizing the multi-particle entanglement.

For a multi-particle system, the characterization of its
entanglement for a pure state usually starts with certain
canonical form of its wave function. In this study, we construct
the standard form of an arbitrary wave function for three
two-state identical bosons. We further characterize its
correlation and squeezing properties based on the standard form.
This paper is organized as follows. In Sec. II, we survey the
important results on the standard form of a pure state wave
function of three distinguishable qubits. We then present our
result for the case of three bosons in two modes in Sec. III,
which is followed by a detailed discussion of entanglement types,
entanglement measures, and spin squeezing properties, respectively
in Sec. IV, Sec. V, and Sec. VI. Finally we discuss the
relationship between spin squeezing and pairwise entanglement in
our system and conclude with a summary.

\section{The standard form of an arbitrary pure state for three qubits}

The Schmidt decomposition does not exist for a three particle pure
state, as proven by A. Peres some time ago \cite{cp5}. We now know
that at least five nonlocal parameters are needed to completely
specify the LU equivalent types of three qubits \cite{cp8}. For
example as was done by Linden {\it et al.} \cite{cp8} using the
method of group theory, the following standard form for three
qubits can be derived:
\begin{eqnarray}
|\psi\rangle_{ABC}&=&\sqrt{\lambda}\,|0\rangle
\left(a|00\rangle+\sqrt{1-a^2}|11\rangle\right)\nonumber\\
& &+\sqrt{1-\lambda}\,|1\rangle
\left[\gamma\left(\sqrt{1-a^2}|00\rangle-a|11\rangle\right)\right.\nonumber\\
& &+f|01\rangle+g|10\rangle\Big],
\end{eqnarray}
where $a$ and $f$ are real numbers and
$\gamma=(1-f^2-|g|^2)^{1/2}$. We note this form is a superposition
of six orthonormal basis states.  In fact, it is found that
 at least five product orthonormal basis
states are needed to express an arbitrary state of three qubits.
This is called the generalized Schmidt form of the canonical form
by Acin {\it et al.} \cite{cp4}, later Acin {\it et al.}
\cite{cp10} further discovered the following ``least
representation", a superposition of five orthonormal basis states
\begin{eqnarray}
\lambda_0|000\rangle+\lambda_1e^{i\phi}|100\rangle+\lambda_2|101\rangle
+\lambda_3|110\rangle+\lambda_4|111\rangle, \label{as}
\end{eqnarray}
where $0\leq\phi\leq\pi$. This form is also called the
``generalized Schmidt decomposition" \cite{cp10}.

If the LU equivalence is used to characterize a three qubit
system, infinitely many different types of entanglement are needed
due to the different values of the five parameters in Eq.
(\ref{as}). To reduce the entanglement types, Bennett {\it et al.}
\cite{cp6} introduced the concept of SLOCC (stochastic LOCC)
reducible. Dur {\it et al.} \cite{cp14} further found that there
is only two SLOCC inequivalent types of entanglement for three
qubits \cite{cp14}: the GHZ type
\begin{eqnarray}
|{\rm GHZ}\rangle={1\over \sqrt{2}}(|000\rangle+|111\rangle),
\end{eqnarray}
and the $W$ type
\begin{eqnarray}
|W\rangle={1\over \sqrt{3}}(|001\rangle+|010\rangle+|100\rangle),
\end{eqnarray}
in contrast to what was known earlier for a system of two parties,
where only one type of entanglement, \textit{i.e.} the EPR type
\cite{cp15}, characterizes all entangled states.

\section{The standard form of an arbitrary pure state for three two-state bosons}
For three identical two-state bosons and
in the first quantization representation,
the general form of its wave function reads
\begin{eqnarray}
|\psi\rangle&=&a|000\rangle+b(|100\rangle+|010\rangle+|001\rangle)\nonumber\\
&&+c(|011\rangle+|101\rangle+|110\rangle)+d|111\rangle.
\end{eqnarray}
After a single particle basis transformation
\begin{eqnarray}
|0\rangle &\rightarrow& \alpha|0\rangle+\beta|1\rangle,\nonumber\\
|1\rangle &\rightarrow& -\beta^{*}|0\rangle+\alpha^{*}|1\rangle,
\label{st}
\end{eqnarray}
the coefficients of transformed basis $|000\rangle$, $|011\rangle$, $|100\rangle$,
and $|111\rangle$ become
\begin{eqnarray}
|000\rangle&:&\
a\alpha^3-d\beta^{*3}-3b\beta^{*}\alpha^2+3c\alpha\beta^{*2},\nonumber\\
|111\rangle&:&\
a\beta^3+d\alpha^{*3}+3b\alpha^{*}\beta^2+3c\beta\alpha^{*2},\nonumber\\
|011\rangle&:&\
a\alpha\beta^2-d\alpha^{*}\beta^{*2}-b\beta^{*}\beta^{2}+c\alpha\alpha^{*2}\nonumber\\
&&+2b\alpha\alpha^{*}\beta-2c\beta\beta^{*}\alpha^{*},\nonumber\\
|100\rangle&:&\
a\beta\alpha^2+d\alpha^{*}\beta^{*2}+b\alpha^*\alpha^2+c\beta\beta^{*2}\nonumber\\
&&-2b\beta\beta^{*}\alpha-2c\alpha^{*}\alpha\beta^{*},
\end{eqnarray}

As proven in Appendix A, we find that the following proposition
holds.

\textbf{Proposition 1}: By properly choosing $\alpha$ and $\beta$
we can make any one of the above four coefficients zero.

Proposition 1 leads to the following direct corollary
with properly chosen phase factors for $|0\rangle$ and $|1\rangle$:

\textbf{Corollary 1}: The wave function of three identical bosons
in two modes can be written in the standard form
\begin{eqnarray}
|\psi\rangle=r|000\rangle+s(|100\rangle+|010\rangle+|001\rangle)+t|111\rangle,
\label{3stand}
\end{eqnarray}
with $r$ and $t$ real.

Our results in the next three sections will be based on this
standard form.

\section{Entanglement types}
As discussed earlier three distinguishable qubits can be entangled
in two different ways \cite{cp14}, denoted by a pure state wave
function of the GHZ or the W type. For three identical bosons, we
give similar definitions for the two different types as in the
following.

\textbf{Definition 1}: Three two-state bosons are GHZ type
entangled if its wave function can be written as
\begin{eqnarray}
|\psi\rangle=|\alpha\alpha\alpha\rangle+|\beta\beta\beta\rangle,
\end{eqnarray}
under appropriate single particle transformations, where
$|\alpha\rangle$ and $|\beta\rangle$ are linear independent
but need not be orthogonal and orthonormal.

\textbf{Definition 2}: Three two-state bosons are W type entangled
iff the wave function can be written as
\begin{eqnarray}
|\psi\rangle=|{\alpha}{\beta}{\beta}\rangle
+|{\beta}{\alpha}{\beta}\rangle +|{\beta}{\beta}{\alpha}\rangle
\end{eqnarray}
under appropriate single particle transformations.
$|\alpha\rangle$ and $|\beta\rangle$ are linear independent
but need not be orthogonal and orthonormal.

Using proposition 1, it is straightforward to prove
\textbf{Proposition 2} (see Appendix B): When written in the
standard form of Eq. (\ref{3stand}), three two-state bosons are
GHZ type entangled iff ($r\neq 0$, $t\neq 0$, and $s=0$), or
($r=0$, $t\neq 0$, and $s\neq 0$), or ($r\neq 0$, $t\neq 0$, and
$s\neq 0$); and they are W type entangled iff ($r=0$, $t=0$, and
$s\neq 0$), or ($r\neq 0$, $t=0$, and $s\neq 0$).

It is interesting to note that the parameters $r$ and $t$ are not
symmetric with interchange to the middle term in the standard form
(\ref{3stand}). This observation is consistent with our
proposition that the state is $W$ type entangled iff $s\neq 0$ and
$t=0$. This point can be understood intuitively as the basis
$|000\rangle$ contains two $|0\rangle$s, and is closer to the $W$
state defined here than the basis $|111\rangle$, thus it is
reasonable that only the state $t=0$ is $W$ entangled.

\begin{figure}[htbp]
\includegraphics[width=3.25in]{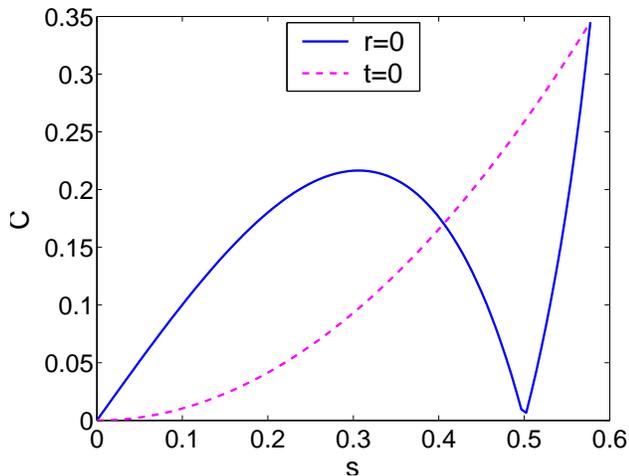}
\caption{The concurrences for a pure state of three bosons
$|\psi\rangle$ as in Eq. (\ref{sf}) for r=0 (blue solid line) or
t=0 (magenta dashed line).} \label{fig1}
\end{figure}

\section{Entanglement measures}
The next task is to measure the entanglement of an arbitrary pure
state of three bosons in two modes. As is well known, the
concurrence $\mathcal{C}$ and the quantity $\tau$, introduced by
Wootters {\it et al.} \cite{woo,cp13} are used to measure pairwise
and ternary entanglement for two and three qubits respectively.
Here we will discuss these entanglement measures for our system of
three two-state bosons.

Let us first review the definitions of the concurrence
$\mathcal{C}$ and the quantity $\tau$ for an arbitrary pure state of
three qubits $A$, $B$, and $C$. The concurrence
$\mathcal{C}_{AB}$ is defined as
\begin{eqnarray}
\mathcal{C}_{AB}=\max\{\lambda_1-\lambda_2-\lambda_3-\lambda_4,0\},
\end{eqnarray}
where $\lambda_1$, $\lambda_2$, $\lambda_3$, and $\lambda_4$ are
the square roots of the eigenvalues, in decreasing order, of the
following operator
\begin{eqnarray}
\rho_{AB}(\sigma_y\otimes \sigma_y)
\rho_{AB}^{*}(\sigma_y\otimes\sigma_y),
\end{eqnarray}
with $\rho_{AB}$ the reduced density matrix of qubits $A$ and $B$.
Similarly, one can define the concurrences $\mathcal{C}_{BC}$,
$\mathcal{C}_{AC}$.

Wootters {\it et al.} \cite{cp13} found that
\begin{eqnarray}
\tau_{ABC}&:=&\mathcal{C}_{A(BC)}^2-\mathcal{C}_{AB}^2-\mathcal{C}_{AC}^2\nonumber\\
&=&\mathcal{C}_{B(AC)}^2-\mathcal{C}_{AB}^2-\mathcal{C}_{BC}^2\\
&=&\mathcal{C}_{C(AB)}^2-\mathcal{C}_{AC}^2-\mathcal{C}_{BC}^2,\nonumber
\end{eqnarray}
where $\mathcal{C}_{A(BC)}$, $\mathcal{C}_{B(AC)}$, and
$\mathcal{C}_{C(AB)}$ are concurrences of the pure state
$|\psi\rangle_{ABC}$ with bipartite partitions $A(BC)$, $B(AC)$ and
$C(AB)$.

Before presenting our results on entanglement for a three boson
pure state, we note that although $\mathcal{C}$ and $\tau$ have
been customarily used for three distinguishable particles
\cite{woo,cp13}, they remain valid for the case of three bosons.
This is so because when we construct the decomposition of the
two-qubit density matrix $\rho$ that adopts the minimum average
pre-concurrence $\mathcal{C}$ (and hence the minimal concurrence
of $\rho$), we start from the eigenvalue decomposition of $\rho$
\cite{woo}, which is automatically symmetrized for a three-boson
system. The quantity $\tau$ as defined is also automatically
invariant under exchange of particles.

We now calculate from the standard form (\ref{3stand}) for three
bosons. For convenience, we rewrite Eq. (\ref{3stand}) as
\begin{eqnarray}
|\psi\rangle=r|000\rangle+se^{i\phi}(|100\rangle+|010\rangle+|001\rangle)+t|111\rangle,\
\label{sf}
\end{eqnarray}
where $r$, $s$, $t$, and $\phi$ are all real with three
of them being independent due to normalization.
A direct calculation leads to the following results
\begin{widetext}
\begin{eqnarray}
\mathcal{C}&=&\sqrt{4t^2s^2+2t^2r^2+4s^4+2\sqrt{t^4s^4+t^4r^2s^2-2s^6t^2+s^4t^2r^2
+s^8+2r^2s^3t^3\cos(3 \phi)}}\nonumber\\
&-&\sqrt{4t^2s^2+2t^2r^2+4s^4-2\sqrt{t^4s^4+t^4r^2s^2-2s^6t^2+s^4t^2r^2
+s^8+2r^2s^3t^3\cos(3\phi)}},
\end{eqnarray}
\end{widetext}
and
\begin{eqnarray}
\tau=4|r^2t^2+4ts^3e^{i3\phi}|,
\end{eqnarray}
where the reduced two party density matrices are identical
$\rho_{AB}=\rho_{BC}=\rho_{AC}$, and can be evaluated
in the single particle basis directly, or more generally
from the two particle reduced density matrix of
a general many body system $\sim\rho^{(2)}_{ijkl}
=\langle a_i^\dag a_j^\dag a_k a_l\rangle/2!$.
When $s=0$, we find $\mathcal{C}=0$, i.e. there
is no pairwise entanglement in the state
$r|000\rangle+t|111\rangle$. Nevertheless, there exists
ternary entanglement $\tau=4r^2t^2$.

\begin{figure}[htbp]
\includegraphics[width=3.25in]{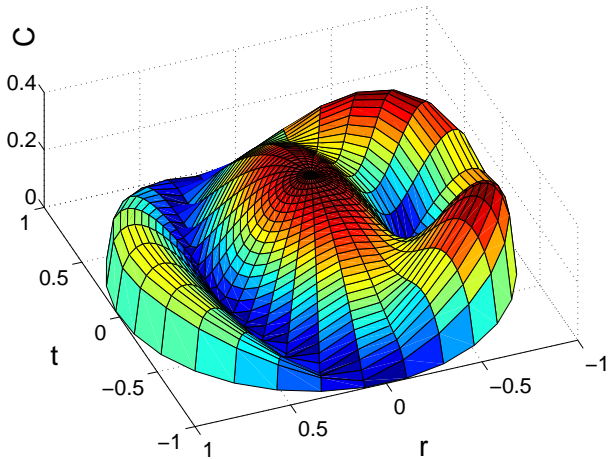}
\caption{The concurrence for a pure state of three bosons
$|\psi\rangle$ as in Eq. (\ref{sf}) when $\phi=0$.} \label{fig2}
\end{figure}

\begin{figure}[htbp]
\includegraphics[width=3.25in]{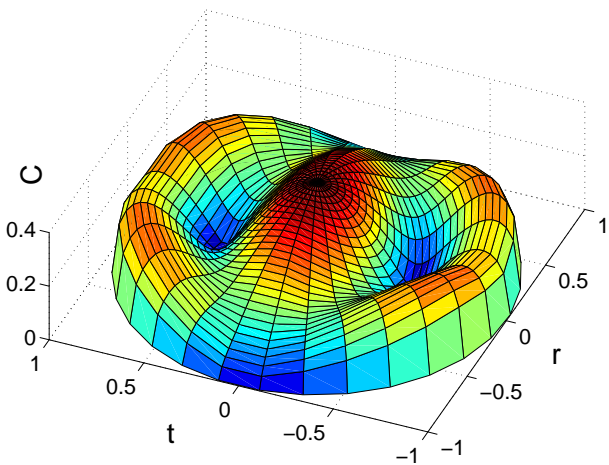}
\caption{The concurrence for a pure state of three bosons
$|\psi\rangle$ as in Eq. (\ref{sf}) when $\phi={\pi}/{2}$. Note it
is symmetric with respect to $t\to -t$.} \label{fig3}
\end{figure}

When $t=0$, i.e., for a $W$ type entangled state, we find
$\mathcal{C}=(\sqrt{6}-\sqrt{2})s^2$ and $\tau=0$.
In this case, the pairwise entanglement
increases with the module of $s$, but there exists
no ternary entanglement. When $r=0$, i.e., for a GHZ type
entangled state, we find
\begin{eqnarray}
\mathcal{C}&=&\sqrt{2}\left|s(\sqrt{3-8s^2}-1)\right|,\\
\tau&=&16\left|ts^3\right|.
\end{eqnarray}
In this case, the concurrence $\mathcal{C}$ vanishes for $s=0$ or
$s=1/2$, where no pairwise entanglement exists. When
$s={\sqrt{3}}/{3}$, the concurrence takes the maximum
$\mathcal{C}={(\sqrt{6}-\sqrt{2})}/{3}$. Another interesting
feature is that there is also a local maximum at $s=
{\sqrt{6}}/{8}$ with a concurrence $\mathcal{C}={\sqrt{3}}/{8}$.
The concurrences for these two special cases of $r=0$ and $t=0$
are shown in Fig. \ref{fig1}. The concurrence $\mathcal{C}$ for a
general pure state is shown in three dimensional graphs, as in
Fig. \ref{fig2} for $\phi=0$ and in Fig. \ref{fig3} for
$\phi={\pi}/{2}$.

\section{Spin squeezing}

Spin squeezing results from quantum correlations between
individual atomic spins \cite{Kit,win}. Recent theoretical
investigations have uncovered that spin squeezing is a sufficient
but not necessary condition for quantum entanglement
\cite{slcp,sor,You2,Zhou,wang,sto,mes}. This has led some effort
to suggest using the spin squeezing parameter as a multi-atomic
entanglement measure \cite{sor}, as has been fully demonstrated in
a two-qubit system \cite{sto}. Wang and Sanders \cite{wang}
illustrated a quantitative relationship between the squeezing
parameter and the concurrence for the even and odd (multiple atom
spin) states, and have further shown that spin squeezing implies
pairwise entanglement for an arbitrary symmetric multi-qubit state
\cite{wang}.

In this section, we investigate the relationship between squeezing
parameter and the pairwise concurrence entanglement measures for
an arbitrary pure state of three two-state bosons. We start from
the standard form Eq. (\ref{sf}) and define the total
``pseudo-spin" for three bosons as
$\vec{S}=(\vec\sigma_1+\vec\sigma_2+\vec\sigma_3)/2$, a direct
calculation then gives
\begin{eqnarray}
\langle\vec{S}\rangle=\left[3rs\cos{\phi},3rs\sin{\phi},\frac{3}{2}(r^2+s^2-t^2)\right].
\end{eqnarray}
It is easy to check that the symmetric three two-state boson
space consists only of the maximum total spin space satisfying
$S^2=(3/2)(3/2+1)\hbar^2$, which implies a geometric Bloch sphere
representation also for the total spin of three two state
bosons. We define the unit vector
$\hat{z}\propto\langle\vec{S}\rangle$ and choose a cartesian
coordinate system with $\hat{x}=(\sin{\phi},-\cos{\phi},0)$,
$\hat{y}\propto[\cos{\phi}(r^2+s^2-t^2),\sin{\phi}(r^2+s^2-t^2),-2rs]$.
This leads to the arbitrary transverse spin direction being
$\vec{n}_\perp=\hat{x}\cos(\theta)+\hat{y}\sin(\theta)$ and
$S_\perp=\vec{S}\cdot\vec{n}_\perp$. The squeezing parameter $\xi$
is defined by
\begin{eqnarray}
\xi=\frac{4}{3}(\Delta S_\perp)_{\rm min}.
\end{eqnarray}
After some tedious calculations, we find
\begin{eqnarray}
\frac{4}{3}(\Delta S_\perp)=A\cos^2{\theta}+B\cos{\theta}\sin{\theta}+C,
\label{eq24}
\end{eqnarray}
with expressions for $A$, $B$, and $C$ given in Appendix C, and
\begin{eqnarray}
{1\over u}=\sqrt{(r^2+s^2-t^2)^2+4r^2s^2}.
\end{eqnarray}

It is reasonably easy to find the minimum of Eq. (\ref{eq24}) since it is a
simple trigonometric function of the form
$(A\cos{2\theta}+B\sin{2\theta})/2+(A/2+C)$, whose minimum can be
found in terms of $2\theta$ and the signs of $A$ and $B$.

We now discuss three important cases for $\xi$ for
an arbitrary pure state of form (\ref{sf}).
First, when $s=0$, we have
$|\psi\rangle=r|000\rangle+t|111\rangle$. In this case, we find
that $\xi$ is always $1$, independent of the values of $r$ and
$t$. This means that these kind of entangled states are never
spin-squeezed.

\begin{figure}[htbp]
\includegraphics[width=3.25in]{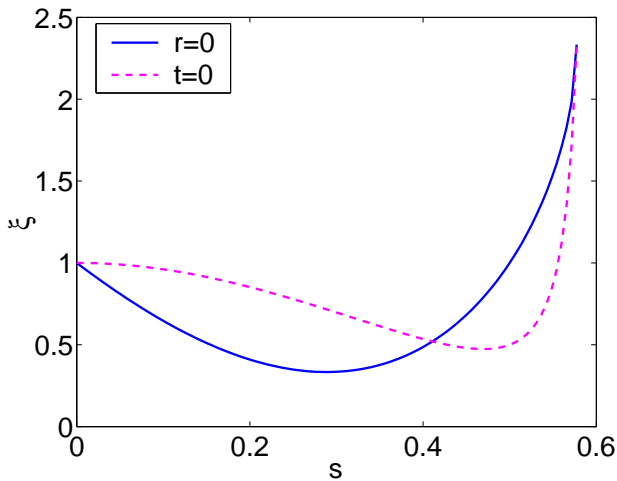}
\caption{The squeezing parameter $\xi$ for a pure state of three
bosons $|\psi\rangle$ as in Eq. (\ref{sf}) with r=0 (blue solid
line) or t=0 (magenta dashed line). } \label{fig4}
\end{figure}

Second, when $t=0$, we find
\begin{eqnarray}
\xi=\frac {1-4s^2+16s^6} {1-8s^4},
\end{eqnarray}
which has one minimum at $s\simeq 0.4694$ with a squeezing
parameter $\xi \simeq 0.4738$. The dependence of $\xi$ as a function
of $s$ is shown in Fig. \ref{fig4} in dashed
line, where $s$ varies from $0$ to $1/\sqrt{3}$. This result is
independent of the value of $\phi$.

\begin{figure}[htbp]
\includegraphics[width=3.25in]{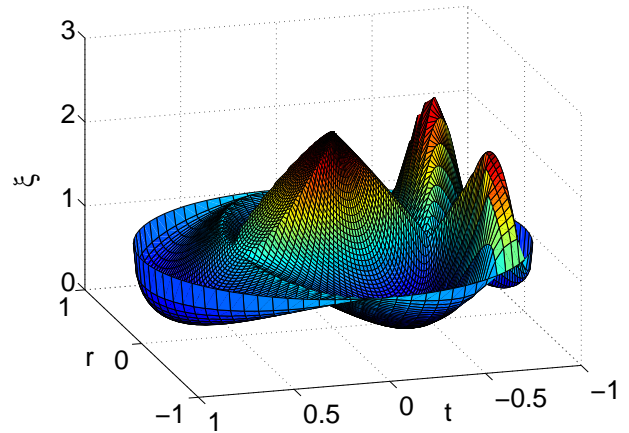}
\caption{The squeezing parameter $\xi$ for a pure state of three
bosons $|\psi\rangle$ as in Eq. (\ref{sf}) with $\phi=0$.}
\label{fig5}
\end{figure}

Third when $r=0$, we find
\begin{eqnarray}
\xi=1+4s^2-4\sqrt{s^2-3s^4},
\end{eqnarray}
which gives $\xi=1$ for $s=0$ and $s={1}/{2}$.
Thus there exists no squeezing in these two states.
The maximum value of the squeezing parameter
is $\xi={7}/{3}$ in this case, corresponding to
$s={\sqrt{3}}/{3}$, i.e. a W state. The minimum value of the squeezing
parameter is $\xi=1/3$ at $s={\sqrt{3}}/{6}$.
The squeezing parameter $\xi$ as a function of $s$ is
plotted in Fig. \ref{fig5}, independent of $\phi$ in this case.

\begin{figure}[htbp]
\includegraphics[width=3.25in]{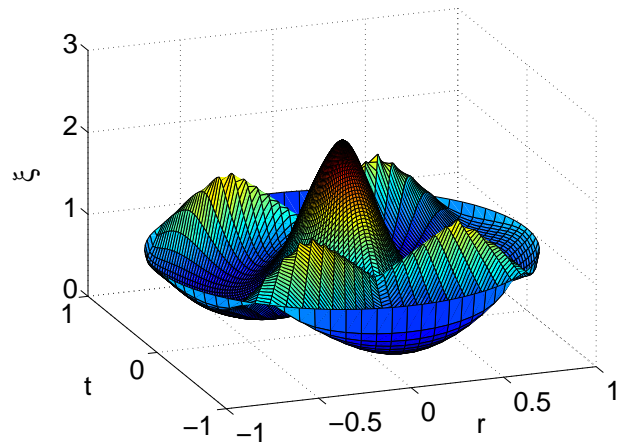}
\caption{The squeezing parameter $\xi$ for a pure state of three
bosons $|\psi\rangle$ as in Eq. (\ref{sf}) with $\phi={\pi}/{2}$.}
\label{fig6}
\end{figure}

Finally, we use two three-dimensional figures to illustrate the
squeezing parameter $\xi$ as a function of $r$ and $t$ in Figs.
\ref{fig5} and \ref{fig6} with both $r$ and $t$ varying from $0$ to
$1$. We have set $\phi=0$ for Fig. \ref{fig5} and $\phi=\pi/2$
for Fig. \ref{fig6}.

\section{Conclusion}
To clarify the relationship between spin squeezing and pairwise
entanglement, let us pay some attention to Figs. \ref{fig1} and
\ref{fig4}. For states of $t=0$, i.e., W-type entangled states,
the concurrence $\mathcal{C}$ is a monotonically increasing
quantity with parameter $s$, while there exists a minimum of the
squeezing parameter $\xi$. Thus, we conclude that for W-type
states the spin squeezing is drastically different from the
pairwise quantum entanglement. For states of $r=0$, i.e., GHZ
tye-entangled states, we find several common features between
pairwise entanglement and spin squeezing: when $s=0$ and $s=1/2$,
neither pairwise entanglement nor spin squeezing exists. For
$s={\sqrt{3}}/{3}$, these two quantities attain the maximum. There
also exists another extreme point in these two quantities.
However, we note that the parameters corresponding to these two
extreme points are not the same. When $s={\sqrt{6}}/{8}$, the
concurrence takes a local maximum value, while the spin squeezing
parameter takes a minimum value when $s={\sqrt{3}}/{6}$. This
different dependence on the parameter $s$ shows that also for
GHZ-type entangled states, the spin squeezing parameter cannot be
regarded as a measure of pairwise entanglement. It is worthy to
point out that on this point our result is consistent with several
previous works \cite{sto,wang,Zhou}. Although for a collection of
special states, there might exists a quantitative relationship
between pairwise entanglement and spin squeezing, these two
properties in general refers to different aspects of multi-party
quantum correlation, and are not simply related to each other.

In summary, we have obtained the canonical form of an arbitrary pure
state for three two-state bosons. Based on this form, we have
classified the entanglement of three identical bosons in two modes
into two types, GHZ and W types, analogues to the case of
three distinguishable qubits \cite{cp14}. We have completely studied
two important entanglement measures, the concurrence $\mathcal{C}$
and the triple entanglement measure $\tau$, and have also
investigated the spin squeezing property of our system by directly
computing the spin squeezing parameter $\xi$. Our results demonstrate
that even for pure states of a system of three bosons in two modes, the spin
squeezing parameter $\xi$ cannot be regarded as an entanglement
measure, in contrast to a system of two particles.

This work is supported by NSF and CNSF.

\appendix
\section{Proof of proposition 1}
Proof: For the coefficient of $|000\rangle$ we prove
that the solution to the equation
\begin{eqnarray}
a\alpha^3-d\beta^{*3}-b\beta^{*}\alpha^2+c\alpha\beta^{*2}=0,
\end{eqnarray}
does exist. Without loss of generality, we assume $\alpha\neq 0$.
Divide the above equation by $\alpha^3$, we get
\begin{eqnarray}
a-d\left(\frac{\beta^{*}}{\alpha}\right)^3
-b\left(\frac{\beta^{*}}{\alpha}\right)+c\left(\frac{\beta^{*}}{\alpha}\right)^2=0.
\label{eq13}
\end{eqnarray}
Of course the solution to Eq. (\ref{eq13}) exits for the variable
${\beta^{*}}/{\alpha}$ from the fundamental theorem of algebra.
Similarly we can prove that the coefficient of $|111\rangle$ can
be eliminated by the single particle transformation Eq.
(\ref{st}).

Now we consider the equation
\begin{eqnarray}
&&a\alpha\beta^2-d\alpha^{*}\beta^{*2}-b\beta^{*}\beta^{2}+c\alpha\alpha^{*2}\nonumber\\
&&+2b\alpha\alpha^{*}\beta-2c\beta\beta^{*}\alpha^{*}=0.
\label{eq5}
\end{eqnarray}
Without loss of generality, we assume $\beta\neq 0$. Let
$z={\alpha}/{\beta^{*}}$ and divide Eq. (\ref{eq5}) by
$\beta^{*}\beta^2$, we get
\begin{eqnarray}
az-dz^{*2}-b+czz^{*2}+2bzz^{*}-2cz^{*}=0.\label{eq:origin}
\end{eqnarray}
Take its complex conjugation, we obtain
\begin{eqnarray}
a^{*}z^{*}-d^{*}z^{2}-b^{*}+c^{*}z^{2}z^{*}+2b^{*}zz^{*}-2c^{*}z=0.\label{eq:comp}
\end{eqnarray}
After eliminating variable $z^{*}$ from Eqs. (\ref{eq:origin}) and
(\ref{eq:comp}), we are left with a fifth order polynomial
equation for the complex variable $z$. According to the
fundamental theorem of Algebra, there exists at least one solution
of equation (\ref{eq:origin}).

A similar procedure can be applied to the coefficient of state
$|100\rangle$. Therefore we complete our proof of proposition 1.

\section{Proof of proposition 2}
Proof:

When $r\neq 0$, $t\neq 0$, and $s=0$, the standard form itself is
just the GHZ type entanglement.

When $r=0$, $t\neq 0$, and $s\neq 0$, we can choose
$|\alpha\rangle=-w|0\rangle+{s|1\rangle}/{2w^2}$ and
$|\beta\rangle=w|0\rangle+{s|1\rangle}/{2w^2}$, where $w$
satisfies $w^6={s^3}/{4t}$. Thus, we get GHZ type entanglement.

When $r\neq 0$, $t\neq 0$, and $s\neq 0$, we choose
$|\alpha\rangle=a|0\rangle+b|1\rangle$ and
$|\beta\rangle=c|0\rangle+d|1\rangle$, where $a =
t^2s^2r/[(tr^2+4s^3)(-t+2u^3)v^2]$, $b=v$,
$c=ru(u^3-t)/s(2u^3-t)$, $d=u$. And $u$ satisfies
$(tr^2+4s^3)u^6+(-t^2r^2-4ts^3)u^3+t^2s^3=0$, $v$ satisfies
$v^3+u^3-t=0$. Thus We will get GHZ type entanglement.

When $r\neq 0,t=0,s\neq 0$, the standard form itself is just the W
type entanglement.

When $r\neq 0,t=0,s\neq 0$, we choose
$|\alpha\rangle={r|0\rangle}/{3}+s|1\rangle$,
$|\beta\rangle=|0\rangle$, and it becomes the W type.

This completes our proof.

\section{Expressions for A, B, and C}
\begin{eqnarray}
A&=&128u^2s^3r^2t\cos^3{\phi}-96u^2s^3r^2t\cos{\phi}-40u^2s^4r^2\nonumber\\
&&-64u^2st^3r^2\cos^3{\phi}-24u^2st^5\cos{\phi}+48u^2s^3t^3\cos{\phi}\nonumber\\
&&-24u^2s^5t\cos\phi-64u^2s^3t^3\cos^3\phi+32u^2s^5t\cos^3\phi\nonumber\\
&&+32u^2st^5\cos^3\phi+24u^2s^2r^2t^2+48u^2st^3r^2\cos\phi\nonumber\\
&&-24u^2sr^4t\cos\phi+32u^2sr^4t\cos^3\phi-8u^2s^2r^4,\\
B&=&-32u t^3 s \cos^2\!{\phi}\sin{\phi} +32 u  s^3 t \cos^2\!{\phi}\sin{\phi}\nonumber\\
&&+32u r^2s t\cos^2\!{\phi}\sin{\phi}-8u r^2 s t\sin{\phi} \nonumber\\
&&-8 u s^3 t\sin{\phi} +8 u t^3 s\sin{\phi} ,\\
C&=&1+4s^2+12st\cos{\phi}-16st\cos^3\!{\phi},
\end{eqnarray}


\begin{thebibliography}{}
\bibitem{slm} J.  Schliemann,  D.  Loss, and A.  H.  MacDonald,  Phys.  Rev. B {\bf 63},
085311 (2001).

\bibitem{You} L. You, Phys. Rev. Lett. \textbf{90}, 030402 (2003).

\bibitem{slcp} A. Sorensen, L. M. Duan, J. I. Cirac, and P. Zoller, Nature
(London) \textbf{409}, 63 (2001).

\bibitem{zzx} B. Zeng, H. Zhai, and Z. Xu, Phys. Rev. A \textbf{66},
042324 (2002).

\bibitem{sckll} J.  Schliemann,  J.  I.  Cirac,  M.  Kus,  M.  Lewenstein, and D.
Loss,  Phys.  Rev.  A {\bf 64},  022303 (2001).

\bibitem{py} R.  Paskauskas and L.  You,  Phys.  Rev.  A {\bf 64},  042310 (2001).

\bibitem{lbll} Y.  S.  Li,  B.  Zeng,  X.  S.  Liu, and G.  L.  Long,  Phys.  Rev.  A {\bf 64},
054302 (2001).

\bibitem{gf} J.  R.  Gittings and A.  J.  Fisher,  quant-ph/0202051.

\bibitem{esbl} K.  Eckert,  J.  Schliemann,  D.  Bru$\beta$,  and M.  Lewenstein,
Annals of Physics (New York) {\bf 299}, 88 (2002).

\bibitem{wiseman}
H. M. Wiseman and J. A. Vaccaro,
Phys. Rev. Lett. {\bf 91}, 097902 (2003).

\bibitem{cp4} A. Acin, A. Andrianov, E. Jane, and R. Tarrach,
J. Phys. A {\bf 34}, 6725 (2001).

\bibitem{cp5} A. Peres, Phys. Lett. A \textbf{202}, 16 (1995).

\bibitem{cp6} C. H. Bennett, S. Popescu, D. Rohrlich, J. A. Smolin, and A. V. Thapliyal,
Phys. Rev. A {\bf 63}, 012307 (2001).

\bibitem{cp8}  N. Linden and S. Popescu,
Fortsch. Phys. \textbf{46}, 567 (1998).

\bibitem{cp10}A. Acin, A. Andrianov, L. Costa, E. Jane, J.I. Latorre, and R. Tarrach,
Phys. Rev. Lett. \textbf{85}, 1560 (2000).

\bibitem{cp14} W. Dur, G. Vidal, and J. I. Cirac, Phys. Rev. A \textbf{62},
062314 (2000).

\bibitem{cp15} C. H. Bennett, H. J. Bernstein, S. Popescu, and B. Schumacher,
Phys. Rev. A \textbf{53}, 2046 (1996).






\bibitem{woo} W. K. Wootters, Phys. Rev. Lett. {\bf 80}, 2245 (1998).

\bibitem{cp13} V. Coffman, J. Kundu, and W. K. Wootters, Phys. Rev. A {\bf 61}, 052306 (2000).


\bibitem{Kit} M. Kitagawa and M. Ueda, Phys. Rev. A \textbf{47}, 5138
(1993).

\bibitem{win} D. J. Wineland, J.J. Bollinger, W. M. Itano, and
D.J. Heinzen, Phys. Rev. A \textbf{50}, 67 (1994).

\bibitem{sor} A.S. Sorensen and K. Molmer, Phys. Rev. Lett.
\textbf{86}, 4431 (2001).

\bibitem{sto} J. K. Stockton, J. M. Geremia, A. C. Doherty, and H. Mabuchi, Phys. Rev. A
\textbf{67}, 022112 (2003).

\bibitem{wang} X. Wang and B. C. Sanders, Phys. Rev. A
\textbf{68}, 012101 (2003).

\bibitem{You2} {\"O}. M{\"u}stecaplio{\u g}lu, M. Zhang, and L. You, Phys. Rev. A
\textbf{66}, 033611 (2002).

\bibitem{Zhou} L. Zhou, H.S. Song, and C. Li, J. Opt. \textbf{B}:
Quantum Semiclassical Opt. \textbf{4}, 425 (2002).

\bibitem{mes} A. Messikh, Z. Ficek, and M. R. B. Wahiddin, Phys.
Rev. A \textbf{68}, 064301 (2003).

\end{thebibliography}
\end{document}